\documentclass[runningheads]{llncs}
\usepackage{graphicx}
\usepackage{float}
\usepackage{multirow}
\usepackage{unicode-math}
\usepackage{hyperref}
\usepackage{booktabs} 

\begin{document}
\title{Exploring the Relationship Between Local Election Results and Online Public Opinion in Taiwan: A Case Study of Taitung County}
\titlerunning{Election Results and Online Public Opinion}
%
\author{I-Hsien Ting\inst{1}\orcidID{0000-0002-6587-2438}, Yen-Chih Chiu\inst{2}, Yun-Hsiu Liu\inst{3}\orcidID{0009-0003-6606-3776} and Kazunori Minetaki\inst{4}\orcidID{0000-0003-4023-068X} and Chia-Sung Yen\inst{3}\orcidID{0000-0002-9021-9940}}
\authorrunning{Ting et al.}
\institute{
National University of Kaohsiung, Taiwan\\
\email{iting@nuk.edu.tw}\\
\and
National University of Kaohsiung, Taiwan\\
\email{kenvie0629@gmail.com}\\
\and
National University of Kaohsiung, Taiwan\\
\email{yvonneliu0616@gmail.com}\\
\and
Kindai University, Japan\\
\email{kminetaki@bus.kindai.ac.jp}\\
\and
National Chinyi University of Technology, Taichung City, Taiwan\\
\email{csyen@ncut.edu.tw}
}
\maketitle              
\begin{abstract}

This study examines the relationship between online buzz and local election outcomes in Taiwan, with a focus on Taitung County. As social media becomes a major channel for public discourse, online buzz is increasingly seen as a factor influencing elections. However, its impact on local elections in Taiwan remains underexplored. This research addresses that gap through a comparative analysis of social media data and actual vote shares during the election period. A review of existing literature establishes the study’s framework and highlights the need for empirical investigation in this area.

The findings aim to reveal whether online discussions align with electoral results and to what extent digital sentiment reflects voter behavior. The study also discusses methodological and data limitations that may affect interpretation. Beyond its academic value, the research offers practical insights into how online buzz can inform campaign strategies and enhance election predictions. By analyzing the Taitung County case, this study contributes to a deeper understanding of the role of online discourse in Taiwan’s local elections and offers a foundation for future research in the field.

\end{abstract}
\section{Introduction}

In recent years, with the continuous advancement and widespread adoption of internet technologies, the internet has become one of the primary channels for political discourse and information dissemination. Within this virtual realm, online volume—referring to the quantity of discussions surrounding specific political events or candidates—has gradually emerged as a significant indicator in political analysis and electoral studies. The discourse and opinions reflected in online volume have the potential to influence voters' political attitudes and voting behavior. As such, understanding the relationship between online volume and electoral outcomes has become a focal point of interest in both academic and political spheres.

Taiwan, characterized by its political pluralism and high level of information accessibility, offers a unique context in which local elections constitute a vital component of the democratic system. These elections not only attract considerable attention from domestic and international scholars but also generate widespread engagement and discussion in online spaces. However, existing research on the correlation between online volume and local election outcomes in Taiwan remains limited, particularly in terms of studies employing data comparison methodologies for in-depth analysis. This study seeks to address this research gap by investigating the relationship between online volume and the results of Taiwan's local elections, utilizing comparative data analysis to provide a more comprehensive understanding of their interconnection.

The primary objective of this study is to investigate the relationship between online volume and the outcomes of local elections in Taiwan. Specifically, the research aims to: (1) analyze the correlation between the number of votes received by local election candidates and the volume of online discussions related to these elections during the corresponding periods; (2) examine the extent to which online volume influences electoral outcomes, as well as the underlying mechanisms through which such influence may occur; and (3) compare variations in the relationship between online volume and electoral results across different regions, time periods, and political contexts. Through these analytical dimensions, the study seeks to provide a more nuanced understanding of the interaction between digital discourse and electoral behavior in Taiwan, thereby contributing novel insights to the fields of political science, communication studies, and the broader social sciences.

This study carries significant theoretical and practical implications. From a theoretical perspective, an in-depth exploration of the relationship between online volume and local election outcomes enhances our understanding of the internet’s role in shaping political behavior and contributes to the expansion of theoretical frameworks in political communication and digital media research. Practically, the findings of this study offer valuable guidance for political candidates and campaign strategists, enabling them to more effectively leverage online platforms in the planning and execution of election campaigns. By shedding light on the dynamics between digital engagement and electoral success, this research supports efforts to improve campaign strategies and increase competitiveness within Taiwan’s democratic processes.

\section{Literature Review}
\subsection{Social Networks and Social Metrics}

Social Network Analysis (SNA) has emerged as a valuable methodological framework for examining public opinion in electoral contexts, particularly through the lens of social interactions and influence patterns within digital platforms. Originally rooted in sociology, SNA provides tools for observing and analyzing the structure and dynamics of social relationships, allowing researchers to uncover the roles and influence of individual actors within a network \cite{b3}. Its conceptual foundation and analytical techniques have been widely adopted across multiple disciplines, including information technology, management, and political science \cite{b4} \cite{b5} \cite{b6}. The fundamental metrics of SNA—such as centrality, betweenness, closeness, clustering coefficient, and structural holes—enable scholars to identify opinion leaders, track information flow, and assess the influence of network position on public discourse. In the context of elections, these metrics help elucidate how political opinions are disseminated, amplified, or silenced across social media, thereby offering critical insights into voter behavior and the formation of public sentiment \cite{b7}.

With the exponential growth of social networking platforms, vast amounts of user-generated data have become available for analysis. This surge in data has attracted the attention of researchers across disciplines, particularly in the fields of information technology and communication. However, the sheer volume of data presents significant challenges for efficient analysis. Opinion mining—also known as sentiment analysis—has emerged as a pivotal solution, enabling the automated extraction of user sentiments from unstructured text. This technique focuses not only on identifying sentiment-laden expressions but also on transforming qualitative data into structured insights. Methods such as corpus-based and thesaurus-based opinion mining allow researchers to classify emotional tone, detect spam, and even forecast political trends based on social media discourse. Despite the widespread use of sentiment analysis, critical sociological dynamics, such as the Spiral of Silence (SOS) effect, have often been overlooked. Ting (2016) addressed this gap by proposing a framework to detect the SOS effect on social media, demonstrating that individuals may refrain from expressing dissenting opinions when perceiving their views as minority-held, thus impacting the reliability of sentiment-based predictions \cite{b1}.

In parallel, the study of social network structures and metrics has deepened our understanding of user behavior within digital communities. Oerez and Ting (2022) examined how structural holes—gaps between non-redundant contacts in a network—can confer strategic advantages to individuals positioned within them \cite{b2}. Their findings revealed that the sustainability of such advantageous positions is strongly influenced by neighborhood similarity, offering nuanced insights into how social capital is formed and maintained in online environments. These structural characteristics are crucial in interpreting opinion dynamics, as users in structurally favorable positions may exert disproportionate influence on discourse, shaping visible sentiment trends. The integration of network theory with sentiment analysis thus provides a more comprehensive framework for understanding opinion formation and diffusion in social media contexts. Together, these studies underscore the necessity of incorporating both network metrics and sociopsychological effects, such as the SOS, to enhance the accuracy and depth of social media analytics.

\subsection{Elections and Online Public Opinion}
The intersection of social media and political behavior has become a critical area of scholarly inquiry, particularly in understanding how digital platforms influence electoral dynamics. Gil de Zúñiga and Valenzuela (2011) conducted a comprehensive cross-national study that demonstrated a significant positive correlation between social media use, social capital, civic engagement, and political participation. Their findings underscore the role of social media not only as a channel for news consumption but also as a catalyst for democratic participation across various cultural contexts. This research provides compelling evidence that online platforms can mobilize users toward political action, thereby altering traditional models of political engagement. The implications of their work are particularly relevant in the context of local elections, where community-based interactions and digital engagement can meaningfully influence voter sentiment and turnout \cite{b8}.

Conversely, other scholars have urged caution regarding the use of social media data in electoral prediction. Jungherr, Jürgens, and Schoen (2012) offered a critical response to prior studies suggesting that Twitter content could reliably forecast election outcomes. Their critique highlighted methodological shortcomings, such as the representativeness of social media users and the overemphasis on volume-based sentiment analysis. They emphasized that while online discourse reflects certain aspects of public opinion, it is insufficient as a standalone predictive tool without considering broader contextual and demographic factors. Together, these contrasting perspectives illustrate both the potential and the limitations of using online public opinion in electoral studies. They provide a nuanced foundation for examining how digital platforms mediate political expression and voter behavior, informing the current study’s investigation into the relationship between online volume and local election results in Taiwan \cite{b9} \cite{b10}.

\section{Research Design and Data Collection}
\subsection{Research Method}
This study adopts the Data Matching Method as its primary research approach to explore the relationship between local election outcomes and online public opinion in Taiwan. As a quantitative research method, data matching involves comparing datasets from different sources to identify correlations or associations between them. In this context, the study matches official data on local election results with corresponding data on online volume—specifically, the amount of discourse related to candidates or political events on digital platforms. Through this comparison, the research aims to quantitatively assess the extent to which fluctuations in online public opinion are linked to electoral performance, providing empirical evidence to support or challenge assumptions about the predictive value of online discourse.

\subsection{Data Collection}

This study utilizes two primary sources of data to investigate the relationship between online public opinion and the outcomes of Taiwan’s 2024 Taitung County legislative election. First, electoral data were collected from the Central Election Commission (CEC) of Taiwan, which provides official statistics on vote counts, vote shares, and the election status of each candidate. Specifically, the study focuses on the three candidates—Liu Chao-hao (Candidate A), Huang Chien-pin (Candidate B), and Lai Kun-cheng (Candidate C) who contested in Taitung County. The data cover election results by township and include variables such as total votes received and winning margins. After collection, the electoral data underwent a thorough cleaning process to ensure completeness, consistency, and accuracy, which involved validating entries, correcting discrepancies, and preparing the dataset for analysis.

Second, online public opinion data were gathered using OpView (\url{https://www.opview.com.tw/}), a web-based social listening platform. This tool was employed to monitor and analyze the online visibility of the three candidates from December 2022 to December 2023. The dataset includes weekly and monthly volume trends, sentiment scores, platform-specific engagement rankings, and demographic breakdowns by gender and user groups. The sources analyzed include major social media platforms (e.g., Facebook, Twitter, Instagram), online forums (e.g., PTT \url{https://www.ptt.cc/bbs/}, Dcard \url{https://www.dcard.tw/}), and news websites. After collection, the data were processed to remove duplicates, filter out irrelevant discussions, and apply sentiment and topic analysis techniques. This preprocessing ensured that the data accurately reflected election-related public discourse, enabling meaningful comparisons with official electoral outcomes.

\section{Data Analysis and Comparison}

\subsection{Online Public Opinion Analysis}

An analysis of online supporters and commentators across various themes reveals distinct gender distributions for each of the three candidates. Candidate A demonstrates a male support rate of 76\% and a female support rate of 23\%. Candidate B shows a relatively balanced distribution, with 46\% male and 53\% female supporters. Candidate C has a male support rate of 79\% and a female support rate of 20\%.

\begin{figure}[H]
\centering
\includegraphics[width=1\textwidth]{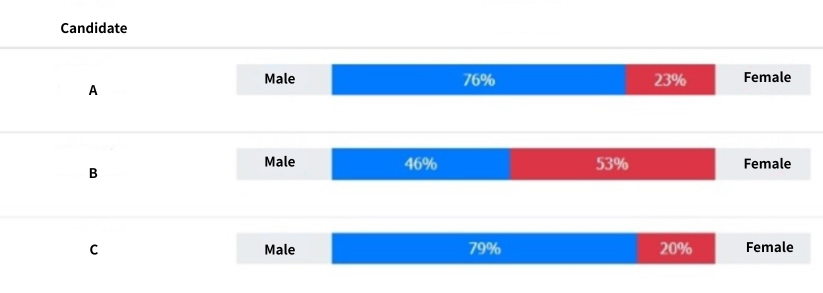}
\caption{Gender Distribution of The Three Candidates} \label{fig1}
\end{figure}

In terms of thematic online discussions, the most frequently occurring topics for Candidate A include news, politics and society, lifestyle sharing, finance and investment, and travel and accommodation. For Candidate B, the dominant themes are politics and society, news, lifestyle sharing, travel and accommodation, and entertainment sharing. Candidate C’s online discourse centers around politics and society, news, finance and investment, meteorology and geography, sports and fitness, arts and culture, as well as travel and accommodation. Notably, Candidate B's online support base is characterized by a more balanced gender distribution, whereas the commentary and engagement for Candidates A and C are predominantly male. The gender distribution of online supporters for each candidate is illustrated in Figure \ref{fig1}.

\begin{figure}[H]
\centering
\includegraphics[width=1\textwidth]{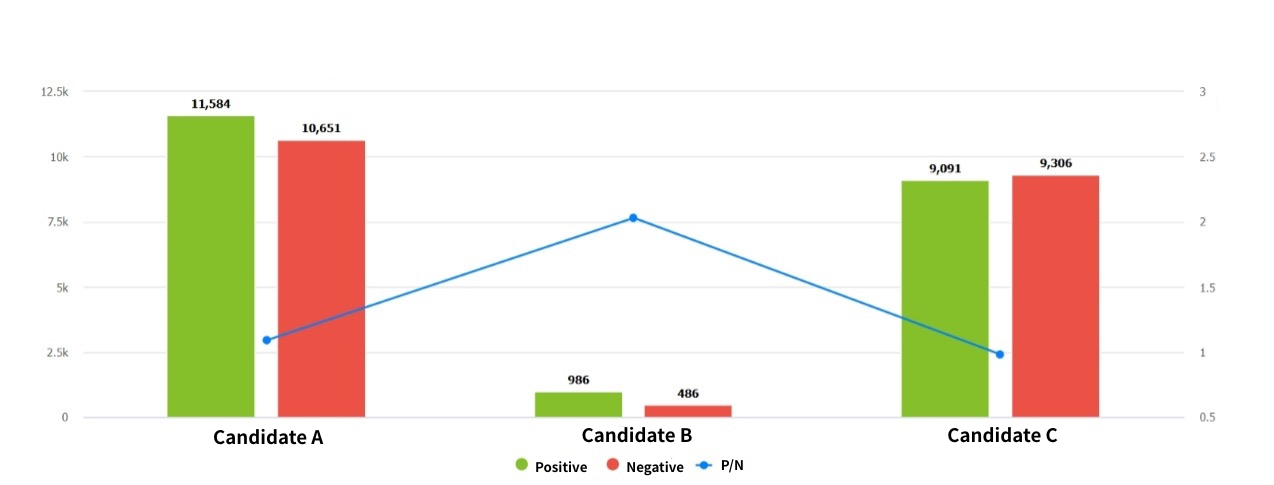}
\caption{Sentimental Analysis of Online Public Opinion} \label{fig2}
\end{figure}

Figure \ref{fig2} presents a sentiment analysis of online discussions across various platforms for the three candidates. Candidate A has a sentiment score of 1.09, with 11,584 positive mentions and 10,651 negative mentions. Candidate B shows a higher sentiment score of 2.03, based on 986 positive mentions and 486 negative mentions. Candidate C has a sentiment score of 0.98, with 9,091 positive mentions and 9,300 negative mentions.
According to the sentiment analysis model, Candidate B has the highest sentiment score, indicating a more favorable public perception. Candidate A's sentiment results are relatively balanced, with a slightly higher number of positive mentions. In contrast, Candidate C has more negative mentions than positive ones, reflecting a less favorable online sentiment.

\begin{figure}[H]
\centering
\includegraphics[width=1\textwidth]{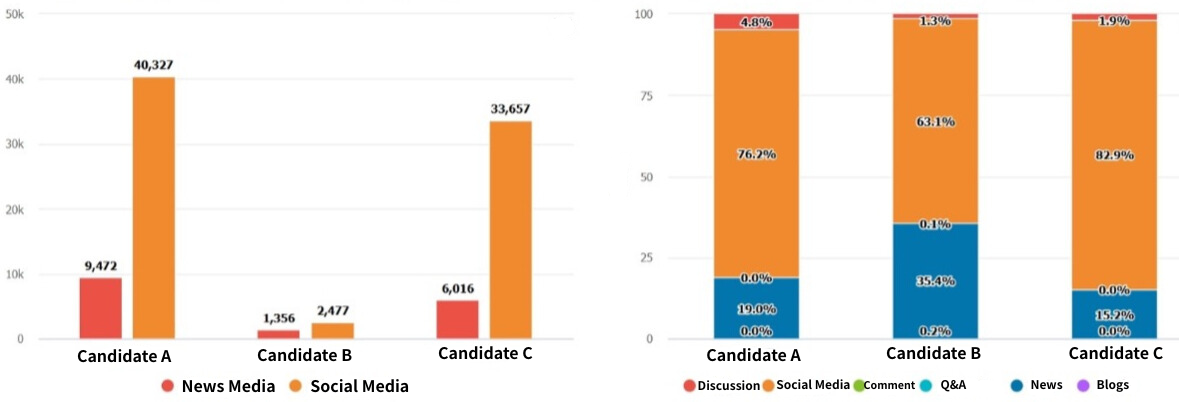}
\caption{The Comparison of Online Public Opinion in Different Platforms} \label{fig3}
\end{figure}

An analysis of online platform distribution reveals significant differences in the digital visibility of the three candidates. Candidate A exhibited the highest overall presence, with 76.20\% of mentions originating from social media, 19.02\% from news outlets, 4.75\% from forums, and a minimal 0.03\% from blogs. Candidate C followed a similar pattern, with 82.89\% of mentions on social media, 15.16\% from news, and limited engagement across other platforms. In contrast, Candidate B had the lowest overall volume, with 63.08\% of mentions from social media, 35.38\% from news, and negligible figures in other categories. This suggests that Candidates A and C maintained stronger engagement across major platforms, particularly in social and news media, whereas Candidate B lagged behind significantly in both visibility and reach.

Further examination of the top 20 websites and channels by topic concentration underscores the differences in communication effectiveness. Candidate A ranked first, generating the highest volume of mentions (30,166) on Facebook fan pages. Candidate C ranked second with 22,070 mentions from Facebook key opinion leaders, indicating substantial influence through KOL networks. In comparison, Candidate B’s most engaged channel—a Facebook fan page with 1,616 mentions—ranked only tenth, further reflecting limited digital traction. These findings indicate that Candidates A and C achieved broader and more impactful digital engagement, while Candidate B's presence remained comparatively weak across all measured dimensions.

\begin{figure}[H]
\centering
\includegraphics[width=1\textwidth]{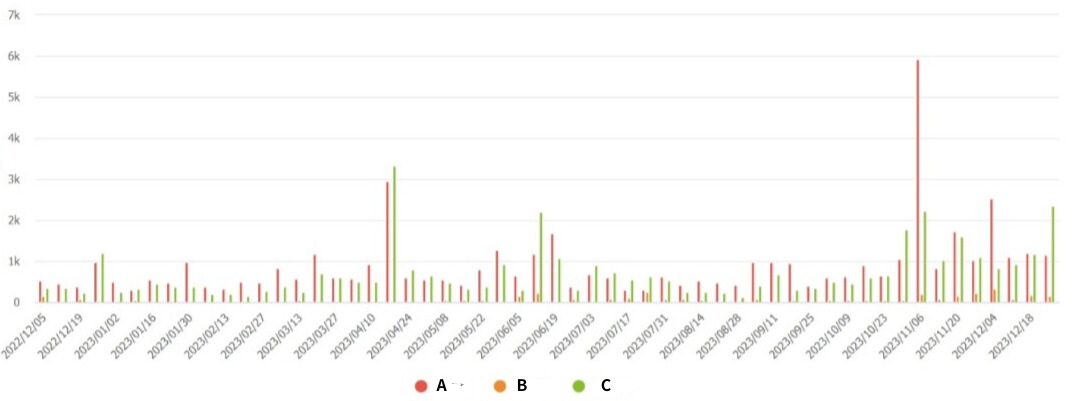}
\caption{Weekly Volume of Online Public Opinion} \label{fig4}
\end{figure}

\begin{figure}[H]
\centering
\includegraphics[width=1\textwidth]{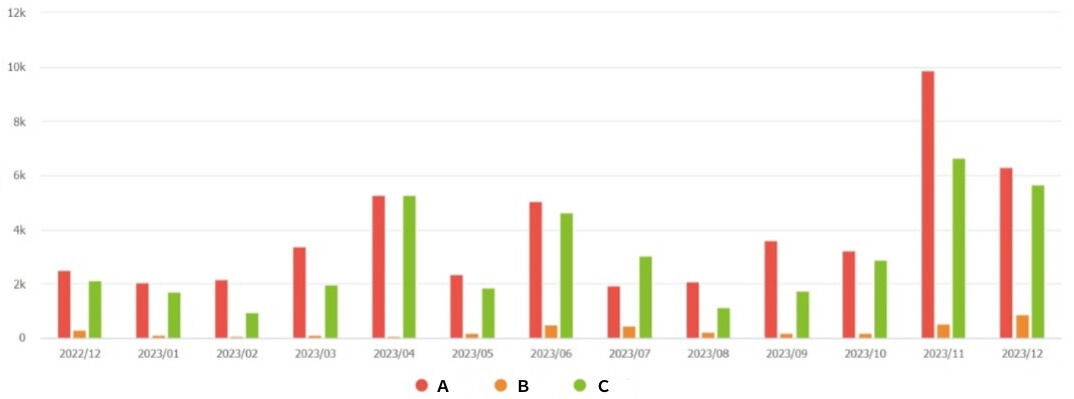}
\caption{Monthly Volume of Online Public Opinion} \label{fig5}
\end{figure}

The analysis of multi-topic volume trends across weekly (Figure \ref{fig4}), monthly (Figure \ref{fig5}), and overall (Figure \ref{fig7}) timelines provides a clear view of how public attention fluctuated over different periods for each candidate. In the weekly analysis, Candidate A experienced peak daily volume on November 6, 2023, with 5,933 mentions, followed by a secondary peak on April 17, 2023, with 2,951 mentions. Candidate B reached their highest single-day volume on December 4, 2023, with 318 mentions, and the second highest on July 24, 2023, with 241 mentions. Candidate C recorded the highest daily volume on April 17, 2023, with 3,336 mentions, and the second highest on December 25, 2023, with 2,361 mentions. These peaks indicate that each candidate experienced surges in public attention at different times, likely reflecting distinct campaign events or media coverage.

In terms of monthly trends, the months of April and November 2023 saw heightened levels of discussion across all candidates, suggesting periods of increased political engagement or significant campaign milestones. The overall volume analysis further underscores the disparity among candidates, with Candidate A receiving the highest total number of mentions at 49,799, followed by Candidate C with 39,673, and Candidate B with only 3,833. This data highlights not only the variation in peak engagement days among the candidates but also emphasizes Candidate A’s consistently stronger presence and Candidate B’s relatively limited visibility throughout the campaign period.

\begin{figure}[H]
\centering
\includegraphics[width=1\textwidth]{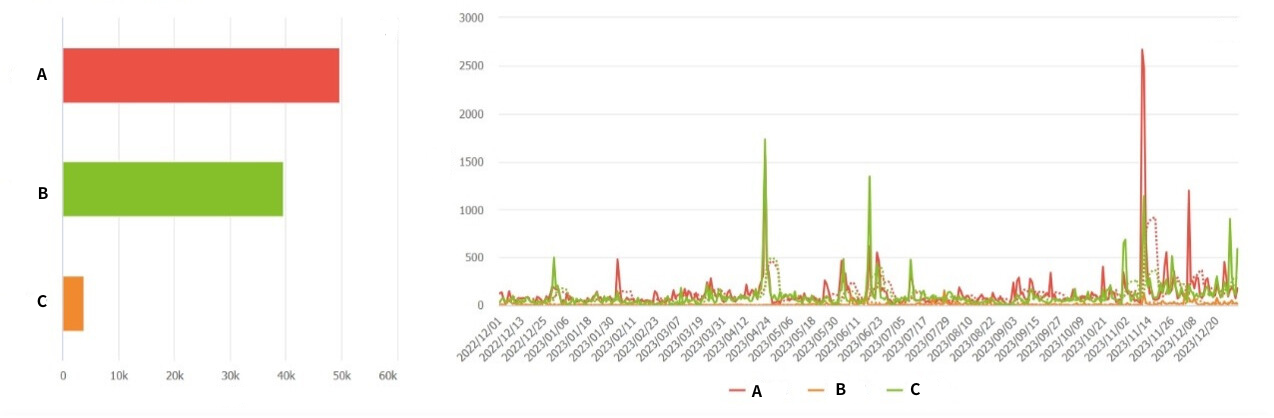}
\caption{Total Volume of Online Public Opinion} \label{fig6}
\end{figure}

\subsection{The Comparison of Actual Vote Rate and Online Public Opinion}
According to data provided by the Central Election Commission (CEC), the official results for the 11th Legislative Yuan election in Taitung County’s Constituency 01 are as follows: Candidate B received 25,778 votes (34.79\%) and was confirmed as the elected legislator. Candidate C followed with 23,420 votes (31.60\%), and Candidate A received 18,744 votes (25.29\%). The total number of registered voters in Taitung County was 115,336, with a voter turnout of 75,221, representing 65.22\%. Of the total ballots cast, 74,103 (98.51\%) were valid votes, while 1,118 (1.49\%) were invalid.

\begin{figure}[H]
\centering
\includegraphics[width=0.62\textwidth]{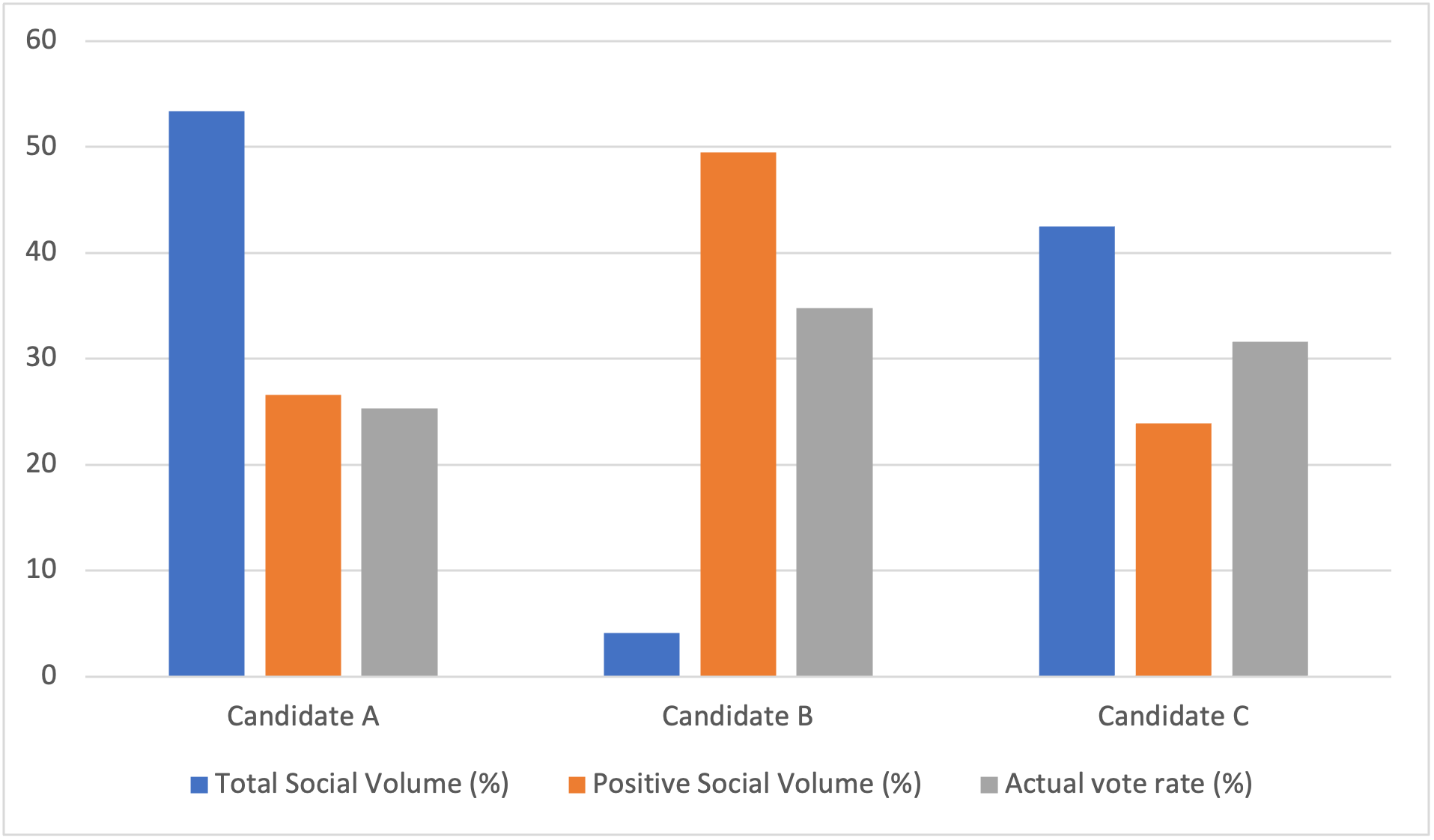}
\caption{The Comparison of Actual Vote Rate and Online Public Opinion} \label{fig7}
\end{figure}

When comparing online presence and sentiment to election outcomes, notable discrepancies emerge. Candidate A garnered the highest volume of online mentions (49,799), yet received the lowest vote share (25.29\%). Their sentiment score was 1.09, ranking second among the three candidates. In contrast, Candidate B had the lowest online visibility with only 3,833 mentions but achieved the highest sentiment score at 2.03, and ultimately won the election with the highest vote share (34.79\%). Candidate C ranked second in both online mentions (39,673) and vote share (31.60\%), but had the lowest sentiment score at 0.98, indicating a higher proportion of negative sentiment. These findings suggest that online volume alone is not a reliable predictor of electoral success; rather, sentiment quality and real-world campaign effectiveness play a more significant role.

\section{Conclusion and Suggestions}
This study employed data triangulation methods to investigate the relationship between online discourse volume and electoral outcomes in local elections in Taiwan. Contrary to initial expectations, the findings revealed an inverse relationship between online attention and actual voting results. While it was hypothesized that greater online presence would correlate positively with electoral success, the analysis demonstrated that high online visibility does not necessarily translate into more votes—particularly in small, localized constituencies. Instead, it is the quality and sentiment of online engagement, rather than the quantity, that appears to play a more decisive role in influencing voter behavior.

These findings highlight the need to reassess certain assumptions often made in political communication research, particularly those related to the impact of social media. The results underscore the importance of considering other influencing factors—such as grassroots campaigning, traditional media exposure, and candidate credibility within the local context—that may outweigh digital visibility. This unexpected outcome opens up valuable opportunities for further inquiry into the nuanced dynamics between online popularity and offline electoral performance.

Future research should adopt a more holistic approach by incorporating social, cultural, and political variables to better understand the complex interplay between digital presence and election outcomes. Comparative studies across different regions and electoral cycles would also shed light on temporal and spatial variations in these dynamics. Moreover, interdisciplinary collaboration—integrating perspectives from political science, communication studies, and sociology—could offer a more comprehensive framework for analyzing electoral behavior in the digital age. While our findings challenge conventional expectations, they pave the way for deeper exploration into the evolving relationship between the internet and democratic processes, with the ultimate goal of enhancing electoral fairness, transparency, and public engagement.

%
%
%
%

\end{document}